\documentclass[%
 reprint,
 amsmath,amssymb,
 aip,
 jcp,
]{revtex4-1}

\usepackage{amsmath}
\usepackage{graphicx}
\usepackage{dcolumn}% Align table columns on decimal point
\usepackage{bm}% bold math
\usepackage{epsfig,color,xspace,multirow,xr,bbold}
\usepackage[all]{xy}
\usepackage{setspace}
\usepackage{url}
\setcitestyle{numbers,square}

\usepackage{hyperref} % Should come last

\hypersetup{
   colorlinks,
   menucolor=blue,
   linkcolor=black,
   citecolor=blue,
   urlcolor=blue
}

\begin{document}

\title{Efficient potential of mean force calculation from multiscale simulations: solute insertion in a lipid membrane}

\author{Roberto Menichetti}
\email{menichetti@mpip-mainz.mpg.de}
\author{Kurt Kremer}
\author{Tristan Bereau}
\affiliation{Max Planck Institute for Polymer Research, 
  Ackermannweg 10, 55128 Mainz, Germany}

\date{\today}

\begin{abstract}
The determination of potentials of mean force for solute insertion in a membrane by means of all-atom molecular dynamics simulations is often hampered by sampling issues. A multiscale approach to conformational sampling was recently proposed by Bereau and Kremer (2016). It aims at accelerating the sampling of the atomistic conformational space by means of a systematic backmapping of coarse-grained snapshots. In this work, we first analyze the efficiency of this method by comparing its predictions for propanol insertion into a 1,2-Dimyristoyl-sn-glycero-3-phosphocholine membrane (DMPC) against reference atomistic simulations. The method is found to provide accurate results with a gain of one order of magnitude in computational time. We then investigate the role of the coarse-grained representation in affecting the reliability of the method in the case of a 1,2-Dioleoyl-sn-glycero-3-phosphocholine membrane (DOPC). We find that the accuracy of the results is tightly connected
to the presence a good configurational overlap between the coarse-grained and atomistic models---a general requirement when developing multiscale simulation methods.
\end{abstract} 

\maketitle

\section{Introduction}

Investigating the behavior of compounds embedded in membrane environments is of fundamental 
importance for biological and pharmaceutical applications, with a variety of interesting physicochemical phenomena occurring in a large interval of time- and length-scales, ranging from microscopic to meso- and macroscopic.
Just to mention a few examples, in addition to the traditional analysis of the permeation of a specific compound through the membrane, which is crucial for drug-delivery applications and will be the subject of this work \cite{hsu2016molecular}, it has been shown that the presence of interfacial proteins or small molecules can affect the curvature \cite{blood2006direct}, the phase diagram \cite{de2009effect} and reshape the lipid domains of bilayer systems \cite{barnoud2014hydrophobic}, or even modify the kinetics of MscL gating \cite{melo2017high}. 

In probing the thermodynamic partitioning of small molecules in phospholipid bilayers, a central role is played by the potental of mean force $G(z)$, which describes how the free-energy of a compound changes
as a function of its distance $z$ from the membrane midplane. In addition to providing structural resolution about the equilibrium properties of bilayer insertion \cite{bemporad2004permeation,paloncýová2014amphiphilic,maccallum2008distribution}, $G(z)$ is directly linked to the partitioning coefficient, an experimentally measurable quantity that has been systematically analyzed for a large set of compounds \cite{sangster1989octanol,jakobtorweihen2014predicting}. Moreover, $G(z)$ can be 
combined with dynamical properties such as the position dependent diffusion coefficient $D(z)$ in order to gain insights into the compound permeation kinetics, quantified in terms of the permeability coefficient \cite{marrink1994simulation,lee2016simulation}.

An accurate determination of the potential of mean force---more generally, of a free-energy surface---by means of classical atomistic molecular dynamics (MD) simulations is often hampered by sampling issues  \cite{neale2011statistical,neale2016sampling}. These may arise as a consequence of high free-energy barriers separating metastable states along the collective coordinates chosen (the distance $z$ in our case). Many strategies in the form of enhanced-sampling techniques have proven to be very successful in solving this problem, such as umbrella sampling simulations, metadynamics, adaptive biasing force, and generalizations \cite{torrie1977nonphysical,barducci2008well,darve2008adaptive}. 
A second source of sampling errors which is more difficult to identify---and consequently overcome---arises from the so-called hidden free-energy barriers \cite{sugita2000multidimensional,zheng2008random,neale2013accelerating,valsson2016enhancing}. Such barriers are embedded into the potentially large set of degrees of freedom orthogonal to the ones under consideration, and are typically characterized by long relaxation times. Because these variables are not visible along the chosen ones, they can trap the simulation into disguised metastable states, causing a severe undersampling even if enhanced techniques are employed. 

These problems would be absent in the ideal case of infinite sampling. In finite MD simulations, they could be clearly mitigated by  longer simulations \cite{neale2011statistical} and/or increasing the number of (relevant) collective variables sampled \cite{valsson2016enhancing}. 
However, despite the rapid increase in computer power, all-atom (AA) MD simulations are still limited to relatively small systems and short timescales.
This limitation does not specifically apply to drug-membrane systems, but characterizes the \emph{in silico} investigation of soft matter as a whole.

One possible way of gaining access to larger simulation time and length scales is to employ coarse-grained (CG) models, which describe the system at a lower level of resolution.
In these effective models, subsets of atoms are grouped together into elementary units, or beads, with interactions that are tuned in order to reproduce a set of target properties of the original system, either structural or thermodynamic \cite{noid2013perspective}.
Besides reducing the number of degrees of freedom, grouping set of atoms into effective interaction sites usually corresponds to a smoothing of the rough atomistic energy (and consequently free-energy) landscape, thus alleviating the difficulties in an adequate conformational sampling.

Among the variety of CG models proposed in the literature, the MARTINI force field \cite{M-2004,M-2007} 
has proven to be very successful in reproducing the thermodynamic properties of compounds embedded in lipid bilayers \cite{M-2008,bereau2015automated}. MARTINI maps a chemical fragment
onto a bead chosen among a set of 18 distinct types, depending on the fragment's water/octanol partition coefficient, hydrogen bond capability and charge. An automated paradigm for determining the MARTINI representation of small molecules has been recently proposed in Ref.~\cite{bereau2015automated}.

The price to pay in applying such coarse-grained descriptions is the loss of local, chemical detail, which may 
have a significant impact on the thermodynamics of drug-membrane systems. More generally, the use of a single level of resolution---either atomistic or coarse-grained---is often insufficient to comprehensively cover the large variety of physical phenomena 
occurring in soft matter, resulting from a continuous and delicate interplay between different time and length scales.

The necessity of dealing simultaneously with different levels of detail paved the way to the introduction of multiscale simulation approaches, in which several resolutions are concurrently employed in the description of a system, within a single framework. It is beyond the scope of this work to comprehensively discuss such methods, excellent reviews being available in the literature  \cite{peter2010multiscale,pluhackova2015biomembranes}.

However, it is important to stress that the prerequisite of an accurate multiscale approach is the development of reliable coarse-grained models starting from AA ones. In addition to correctly capturing the large scale behavior of the system, this is fundamental when a reconstruction of the high-resolution detail starting from the low-resolution description is needed, a procedure commonly referred to as backmapping \cite{stansfeld2011coarse,wassenaar2014going,zhang2014equilibration}. Indeed, one must ensure that the backmapped AA configurations are truly representative of the equilibrium ensemble of the atomistic system.
This is far from trivial especially in the case of top-down coarse-grained models as MARTINI, which are often parametrized in terms of little---or  no---structural information.

A multiscale approach to conformational sampling (MACS) for determining the potentials of mean force of solute insertion
in a membrane was recently proposed in Ref.~\cite{bereau2016protein} by Bereau and Kremer. The MACS method aims at accelerating the exploration
of the complex atomistic conformational space by means of a systematic backmapping of uncorrelated coarse-grained snapshot onto the atomistic detail, followed by short AA simulations. This procedure partially mitigates the effect of high free-energy barriers that could prevent an adequate sampling when a single, long atomistic simulation is used. 

MACS was first applied to determine the atomistic potential of mean force of the protein backbone unit into a POPC membrane. The convergence properties of the method were then analyzed, including its sensitivity with respect to both the atomistic and coarse-grained models. Morevorer, the predictions of the method in the case of the insertion of a whole WALP16 peptide into a POPC membrane were compared with those obtained by means of long atomistic simulations, finding a reasonable agreement given the complexity of the molecule.

In this work, we extend previous analyses and probe the overall accuracy of the MACS method in assessing the stability of small molecules in a membrane environment, taking propanol as a test case. We first analyze the case of a DMPC membrane, for which 
reference atomistic results are available. We then investigate the role of two different MARTINI coarse-grained representation of the DOPC lipid in the accuracy of the resulting MACS potentials of mean force, providing insights into the delicate relationship occurring between the coarse-grained and atomistic conformational ensembles.

\section{Methods}
Coarse-grained (CG) molecular dynamics simulations were performed in GROMACS \cite{abraham2015gromacs},
and with the MARTINI force field \cite{M-2004,M-2007,M-2008,M-2013}. 
The integration time step was $\delta t=0.02~\tau$, where $\tau$ is the model's natural unit of time,
and we relied on the standard force-field parameters reviewed in Ref.~\cite{de2016martini}.
Sampling from the $NPT$ ensemble at $P=1$ bar and $T=300~$K was achieved by means of 
a Parrinello-Rahman barostat \cite{parrinello1981polymorphic} and a stochastic
v-rescale thermostat \cite{bussi2007canonical}, with coupling constants 
$\tau_\textup{P}=12~\tau$ and $\tau_\textup{T}=\tau$.
In CG simulations,
a membrane of $\approx\!36 \text{ nm}^2$, containing $N_{L}=128$ lipids (64 per layer) 
was generated by means of the INSANE building
tool \cite{wassenaar2015computational} and subsequently minimized, heated up,
and equilibrated. The number of water molecules surrounding the lipids were 
$N_W=2150,1890,1590$ for DMPC, DOPC M4B and M5B, respectively (see Results).
 As usual when using non-polarizable MARTINI water, we added an additional 
 $\approx10\%$ of antifreeze particles in the system \cite{M-2007}.
 
The CG potential of mean force $G(z)$ was determined by means 
of umbrella sampling simulations \cite{torrie1977nonphysical}.
We set 24 harmonic biasing potentials with $k=240$ kcal/mol/$\text{nm}^2$ every
$0.1$ nm along the normal to the bilayer midplane.
In every simulation, two solute molecules were placed in the membrane
in order to increase sampling and alleviate leaflet area
asymmetry \cite{maccallum2008distribution,bereau2014more,jakobtorweihen2014predicting}.
The total production time for each umbrella simulation was
$1.2\cdot10^5\hspace{1pt}\tau$.
Unbiased free-energy profiles were extracted by means of weighted histogram
analysis method \cite{kumar1992weighted,bereau2009optimized,hub2010g_wham}, and
the corresponding errors via bootstrapping \cite{mooney1993bootstrapping}.

The AA potentials of mean force $G(z)$ of propanol in the three membrane environments 
were determined via the MACS method recently proposed in Ref.~\cite{bereau2016protein}.
In practice, we extracted a set of 30 uncorrelated snapshots
from each of the 24 CG trajectories,
and for each phospholipid bilayer investigated.
The configurations were then converted to the CHARMM force field  \cite{klauda2010update} by means of the
backmapping implementation of Wassenaar et al. \cite{wassenaar2014going}.
This method reconstructs the high-resolution 
system (AA) from the low-resolution one (MARTINI beads) through a geometrical projection, 
combining this with a set of energy minimizations and $NVT$ equilibrations. 

Each set of 30 atomistic configurations provided the initial conditions for short umbrella 
sampling simulations, centered around the value of $z$ equal 
to the one of the coarse-grained simulation from which the snapshots were extracted.
We employed the same harmonic constant $k=240$ kcal/mol/$\text{nm}^2$, and
an integration timestep of $\delta t=0.002~\tau$.
We equilibrated each configuration for $t=100~\tau$ at $P=1$ bar and $T=300K$ by means 
of a Berendsen barostat and Langevin stochastic dynamics ($\tau_\textup{P}=5~\tau$, $\tau_\textup{T}=1.0\tau$).
Subsequently, for the production runs we replaced the Berendsen barostat with a Parrinello-Rahman barostat and the
Langevin dynamics with MD and a v-rescale thermostat ($\tau_\textup{T}=0.1\tau$). The total production time amounted to $t=800~\tau$.

Finally, we recombined the outcomes and obtained the AA potentials of mean force
by means of weighted histogram
analysis method \cite{kumar1992weighted,bereau2009optimized,hub2010g_wham}.

\section{Results}
\begin{figure}[htbp]
  \begin{center}
    \includegraphics[width=\linewidth]{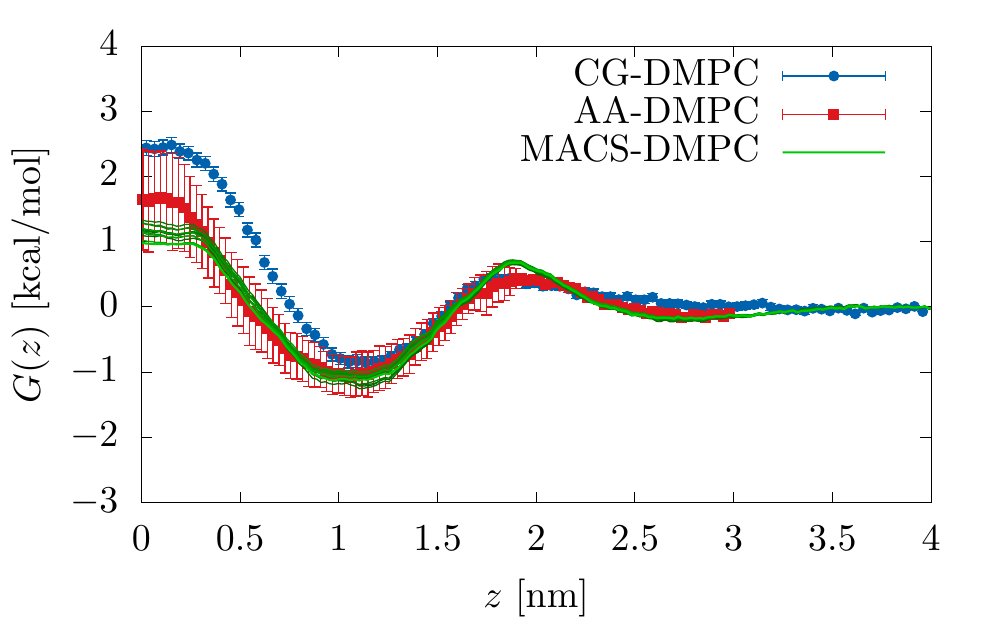}
    \caption{Potentials of mean force for propanol insertion in a DMPC membrane obtained by 
      CG simulations (CG-DMPC), atomistic simulations of $\approx2\mu s$ \cite{jakobtorweihen2014predicting} (AA-DMPC), and the MACS method (MACS-DMPC). Thinner lines correspond to convergence profiles obtained by averaging over an increasing number of short atomistic simulations.}
       \label{fig:dmpc_comp}
  \end{center}
\end{figure}

We first tested the robustness of the MACS method by computing the potential of mean force for the insertion of a propanol molecule into a DMPC membrane. 
The resulting profile is reported in Fig.~\ref{fig:dmpc_comp}, together with its CG counterpart. For further comparison, we also report the curve presented in Ref.~\cite{jakobtorweihen2014predicting}, obtained by means of AA umbrella sampling simulations of the CHARMM force field, with a cumulative production time of $t\approx 2\mu$s (corresponding to $t=2\cdot10^6~\tau$ in reduced units).

The coarse-grained potential of mean force slightly overestimates both atomistic results, the largest deviation being $\approx 1.5$ kcal/mol close to the hydrophobic core of the membrane ($z=0$). Small deviations between the CG and the AA bilayer thickness were compensated with a $\approx0.2$ nm horizontal shift of the CG profile.

The curve we obtain by means of the MACS approach is in striking agreement with the atomistic results of Ref.~\cite{jakobtorweihen2014predicting}, the only significant difference---i.e. outside the error bars---consisting in a small free-energy barrier located around $z\approx 1.9$ nm. Deviations are noticeably smaller than those reported in Ref.~\cite{bereau2016protein} for the WALP16 peptide. This is reasonable, given the complexity of WALP16 compared to propanol.

In Fig.~\ref{fig:dmpc_comp} we also report the profiles obtained by averaging over an increasing number of short AA simulations, $N_{AA}=20,\ldots,30$, in order to investigate the convergence properties of the multiscale method. Fluctuations in the curves are small over the whole $z$ range, showing a slight increase close to the bilayer midplane. This is again not surprising, as this region is well known to be often affected by sampling errors \cite{neale2013accelerating}. 

Our results confirm the efficiency of the MACS method in accelerating the configurational sampling of the atomistic phase space.
We were indeed able to obtain accurate potentials of mean force with a total computational time of $10^4$ CPU hours (including coarse-grained simulations, backmappings and short atomistic simulations), to be compared with an estimate of $10^5$ CPU hours needed for ``traditional'' AA molecular dynamics simulations together with enhanced sampling techniques \cite{carpenter2014method}.
\newline\newline

\begin{figure}[htbp]
  \begin{center}
    \includegraphics[width=\linewidth]{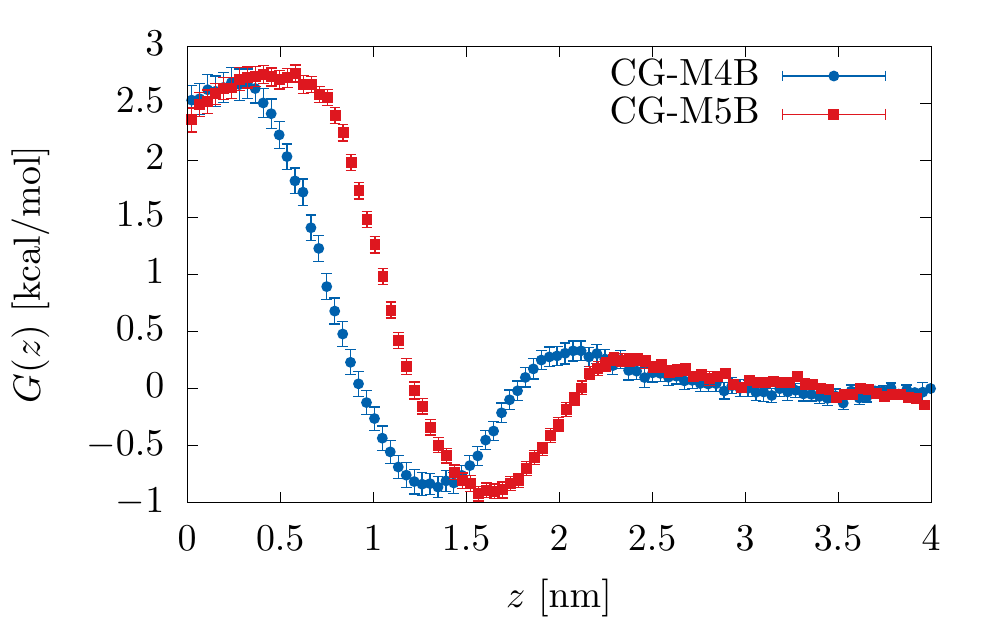}
    \caption{Coarse-grained potentials of mean force for propanol insertion in a DOPC membrane obtained from two different MARTINI representations of the oleoyl terminal chains: 4 beads (CG-M4B) and 5 beads (CG-M5B) models.}
    \label{fig:cg_comp}
  \end{center}
\end{figure}

The continual exchange of information between different levels of resolution introduced in multiscale simulation approaches requires an accurate parametrization of the CG model against the AA detail. 
In our case, this is fundamental for ensuring the presence of a substantial ``configurational overlap" between the two resolutions, so that the backmapping of an equilibrium CG snapshot is representative of the atomistic equilibrium ensemble \emph{at the same thermodynamic state point}. This is the keystone that allowed us to perform very short equilibration runs over the backmapped AA configurations before starting the (short) production simulations. 

However, because the high-to-low resolution mapping is far from unique, distinct effective coarse-grained models can be developed for a given system. 
It is then interesting to test the sensitivity of the method with respect to a change in the CG representation, as this may result in slightly different conformational ensembles when the CG configurations of different models are backmapped onto the same atomistic detail.
Therefore, we computed the potential of mean force for the insertion of propanol into a DOPC membrane by applying MACS to two different MARTINI lipid models that were proposed in the literature. They differ in the number of coarse-grained beads composing each DOPC terminal oleoyl chain: the old model contains 5 beads (M5B), and the updated model 4 beads (M4B). 
The MARTINI building block strategy consists in mapping sets of 4 heavy atoms into a CG bead, so that the 18 carbon atoms composing each DOPC tail falls exactly in between a representation in terms 4 or 5 CG beads. Moreover, the updated model should provide a more realistic phase diagram \cite{wassenaar2015computational}.

  A similar analysis was already performed in Ref.~\cite{bereau2016protein} in the case of leucine in a POPC membrane backmapped to the GROMOS force field, showing relatively small deviations.

\begin{figure}[htbp]
  \begin{center}
    \includegraphics[width=\linewidth]{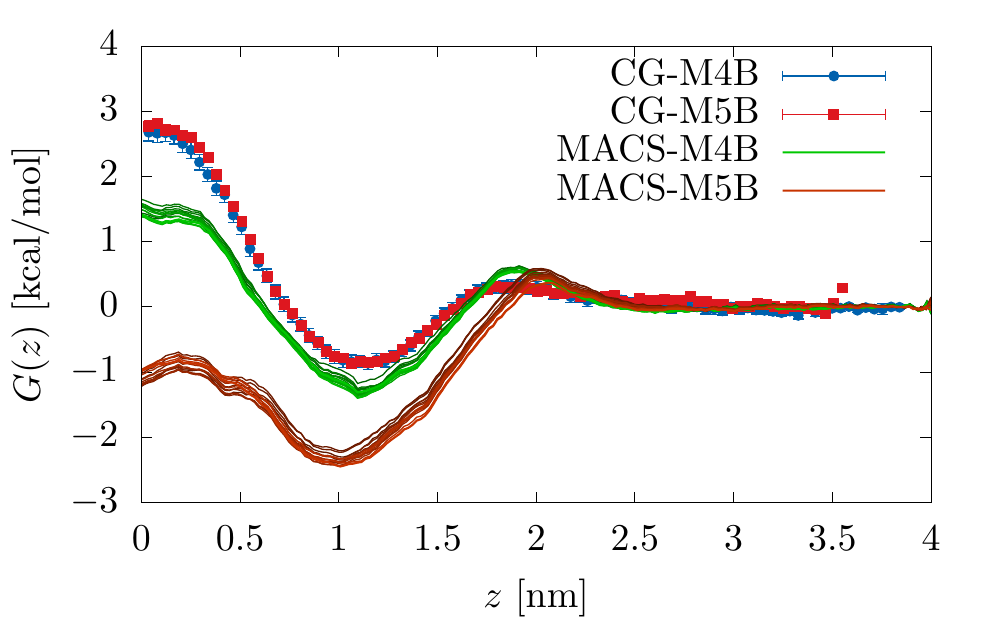}
    \caption{Potentials of mean force of propanol in a DOPC membrane obtained by means of the MARTINI coarse-grained M4B and M5B model (CG-M4B and CG-M5B), and the corresponding MACS profiles. Thinner lines correspond to convergence profiles obtained by averaging over an increasing number of short atomistic simulations.}
    \label{fig:aa_backmap}
  \end{center}
\end{figure}

A comparison of the CG potentials of mean force of the two models in Fig.~\ref{fig:cg_comp} shows
that the two representations are substantially equivalent. The only difference is a relative shift along the reaction coordinate of $\approx 0.3$ nm, which is a consequence of the presence of the longer tails of the M5B model. A shifted version of the two curves is presented in Fig.~\ref{fig:aa_backmap}, showing that they are indeed in perfect agreement.

In Fig.~\ref{fig:aa_backmap}, we report the AA potentials of mean forces of the two models obtained with the MACS method together with their corresponding CG counterparts. The coarse-grained $G(z)$ have been horizontally shifted in order to account for differences in the membrane thicknesses.
The MACS potential of mean force of the M4B model agrees reasonably well with the CG profile, deviations being at most 1 kcal/mol close to the bilayer midplane ($z\approx0$), the same behavior we observed for DMPC. On the other hand, the potential of mean force we obtain with the MACS method on the M5B model is 
not even in qualitative agreement with the other curves. Beside underestimating the free-energy barrier at the bilayer midplane ($\approx 4$ and 3 kcal/mol difference with the coarse-grained and MACS-M4B systems, respectively), the MACS-M5B curve is unable to capture the polar nature of propanol.
These results differ from those presented in Ref.~\cite{bereau2016protein} for the insertion of a leucine side chain in a POPC membrane, where the maximum deviations between analogous curves amounted to $1$ kcal/mol.

The small fluctuations in the convergence profiles presented in Fig.~\ref{fig:aa_backmap}, obtained by averaging over an increasing number of atomistic simulations, $N_{AA}=20,...,30$, show that the difference between the curves does not arise from a sensitive dependence upon the number of small AA simulations employed in the averages.
The source of inaccuracy is then an insufficient equilibration of the backmapped M5B configurations before starting the production simulations. Therefore, this model exhibits a poor conformational overlap with the underlying atomistic system: backmapping the low-resolution system does not generate representative configurations of the equilibrium (Boltzmann) ensemble at the high-resolution level.

\begin{figure}[htbp]
  \begin{center}
    \includegraphics[width=\linewidth]{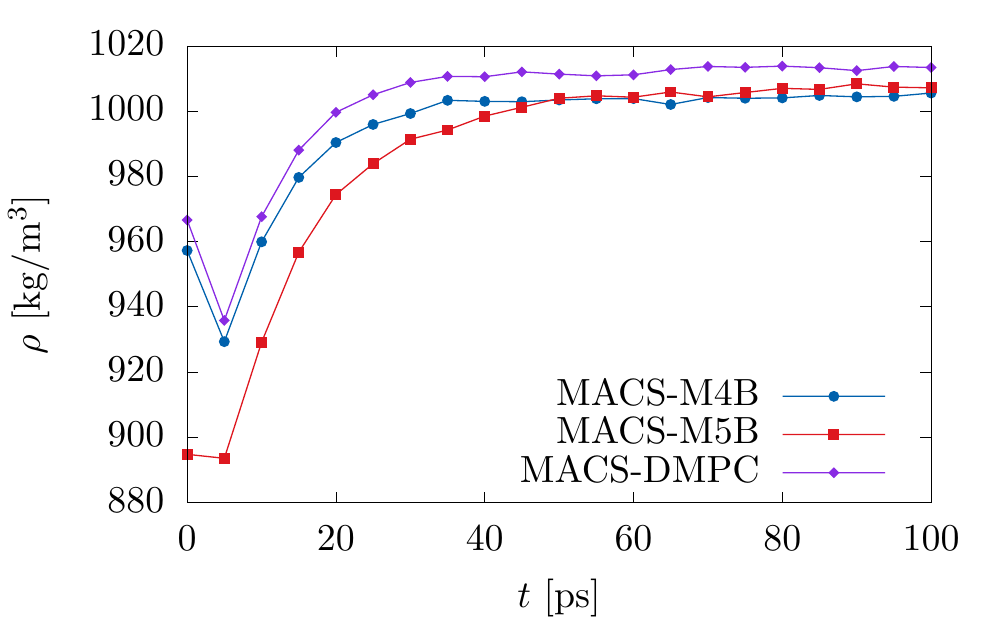}
    \caption{Density plot as a function of time for MACS on the M4B and M5B models for DOPC, and DMPC membranes. Results are shown for one of the short equilibration simulations.}
    \label{fig:dens_plot}
  \end{center}
\end{figure}

We checked this explicitly by analyzing the convergence of different thermodynamic properties along the equilibration simulations in
the two different models. As a representative quantity, in Fig.~\ref{fig:dens_plot} we report density as a function of time.
Both models underestimate the total density at $t=0$. This is not surprising, as it is well 
known how replacing $10\%$ of MARTINI water beads with the slightly bigger antifreeze particles results in a density decrease of roughly $10\%$ \cite{M-2007}.
However, the backmapping of the M5B model shows a lower initial density with respect to the one of the M4B model, and a slightly longer transient time towards convergence ($\approx 65\tau$, compared with $\approx 40\tau$ for M4B). 

It is also interesting to compare the results of the M4B model for DOPC with the case of a DMPC membrane, where MACS succeeds in determining the potential of mean force. The two curves follow the same trend, and have the same time interval for the initial transient regime.

In terms of the simulation box, the density transient shown in Fig.~\ref{fig:dens_plot} corresponds to an observed shrinkage in the total height of the system along the equilibration runs that amounts to $\approx15\%$ in the case of the M5B model, compared with $\approx 5\%$ for the M4B model. 

The deviations are a consequence of the additional beads in the lipid tails of the M5B model. The reason is intuitive: these beads increase the thickness of the bilayer at the coarse-grained level, so that the initial backmapped atomistic configurations present stretched lipid tails that undergo a strong compression in order to reach the atomistic equilibrium density (in addition to the compression arising from antifreeze particles).

This effect becomes drastically more pronounced if one replaces in the equilibration simulations the Langevin dynamics with MD coupled to a v-rescale thermostat. 
The main difference between the two lies in the local nature of the former, so that Langevin dynamics is generally recommended in order to obtain a smooth relaxation of initial configurations with significant deviations from equilibrium.

\begin{figure}[htbp]
  \begin{center}
    \includegraphics[width=\linewidth]{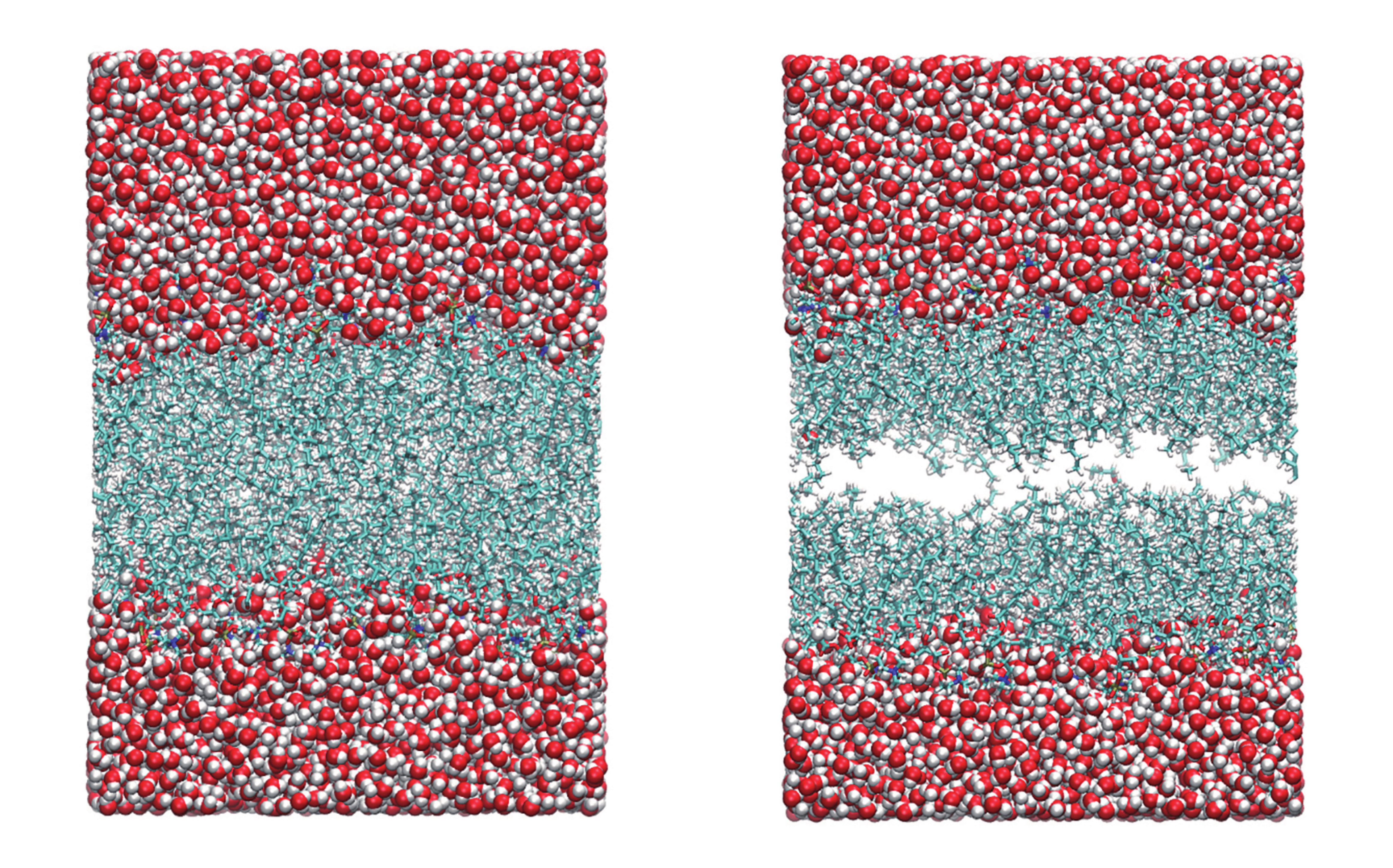}
    \caption{Snapshots of the atomistic configurations obtained from the backmapping of the M5B model during the equilibration simulations for $t=0$ (left) and $t=10\tau$ (right), where an empty slab in the bilayer midplane is generated.}
    \label{fig:slabs}
  \end{center}
\end{figure}

In Fig.~\ref{fig:slabs} we report two atomistic snapshots of a MD-based equilibration simulation starting from a backmapped M5B configuration at $t=0$. At $t=10\tau$ we observe the generation of a vacuum slab between the two lipid layers which arises from the combined relaxation of the initially stretched lipid chains and of the surrounding water molecules. The slab is subsequently reabsorbed via a fast compression of the system in the $z$ direction. The same analysis in the case of the M4B model does not result in the generation of such a slab.

It is then clear how the slightly inaccurate atomistic conformational ensemble generated by the presence of the additional beads in the M5B model requires the system to perform a further relaxation in order to reach equilibrium. Although the corresponding fluctuations seem to have been thermodynamically reabsorbed along the equilibration runs (in both the M4B and M5B models the total density has reached convergence, see Fig.~\ref{fig:dens_plot}), their impact on the microscopic detail can be such that a local equilibration is still far from being reached.
The use of this inadequate configurational ensemble as a starting point for the short production simulations results in the incorrect potential of mean force shown in Fig.~\ref{fig:aa_backmap}.

\section{Discussion}

The accurate determination of potential of mean force for solute insertion in a lipid bilayer from MD simulations is often hampered by the presence of high free-energy barriers, which can cause a severe undersampling even if enhanced techniques are employed \cite{neale2011statistical,neale2016sampling}.
The MACS method proposed in Ref.~\cite{bereau2016protein} aims at accelerating the exploration of the atomistic configurational phase space by performing a systematic backmapping of uncorrelated coarse-grained snapshots onto the atomistic detail, followed by a set of short AA simulations.

In this work, we have critically analyzed the method for propanol insertion into phospholipid DMPC and DOPC bilayers.

We first confirmed the efficiency of MACS by comparing its predictions for the potential of mean force with independent AA simulations \cite{jakobtorweihen2014predicting}. It is important to stress that this method---provided that the CG model presents good configurational overlap with the atomistic system---guarantees accurate results with a computational cost in terms of CPU hours that is one order of magnitude smaller than the one characterizing ``traditional'' AA molecular dynamics simulations. 

We then discussed in detail the role of the coarse-grained representation in affecting the results we obtain via the MACS method. We separately backmapped to the same atomistic detail two different CG models representing DOPC molecules, only differing in the number of beads composing the lipid tails.
This analysis showed that particular attention must be paid in developing reliable CG models, as a substantial configurational overlap with the underlying atomistic detail is required so that the method provides accurate results. 
Indeed, the choice of unoptimal CG representations may cause the backmapped configurations to be not representative of the atomistic equilibrium ensemble. These configurations require a further careful equilibration before being used in production simulations. 
These results are not specific to the MACS method itself, but should be carefully taken into account when performing
multiscale simulation approaches to soft matter in general.

\section*{Acknowledgments}
We thank Joseph F. Rudzinski, Kiran H. Kanekal and Omar Valsson for a critical reading of the manuscript and useful comments.
We acknowledge funding from the Emmy Noether program of the Deutsche Forschungsgemeinschaft (DFG).
\section*{\refname}
\bibliography{biblio} 

\end{document}